\documentclass[a4paper,11pt]{article}
\pdfoutput=1
\usepackage{jheppub}
\RequirePackage{lineno}
\usepackage{overpic}
\usepackage{graphicx}
\usepackage{epsfig}
\usepackage{dcolumn}
\usepackage{bm}
\usepackage{rotating}
\usepackage{multirow}
\usepackage{subfigure}
\usepackage{hhline}
\usepackage{amssymb}
\usepackage{lineno}
\usepackage{hyperref}
\usepackage{mathrsfs}
\usepackage{verbatim}
\usepackage{amsmath}
\usepackage{amssymb}
\usepackage{amsfonts}
\usepackage{amsbsy}
\usepackage{epstopdf}
\hypersetup{colorlinks,linkcolor=blue,anchorcolor=blue,citecolor=blue}
\include{def-com}

\renewcommand\figurename{\rm Figure}
\renewcommand\tablename{\rm Table}
\let\oldequation\equation
\let\oldendequation\endequation

\renewenvironment{equation}
  {\linenomathNonumbers\oldequation}
  {\oldendequation\endlinenomath}

\title{\boldmath Observation of the decay $\psi(3686) \to \Sigma^-\bar\Sigma^+$ and measurement of its angular distribution}
\collaborationImg{\includegraphics[width=.12\textwidth,origin=c,angle=90]{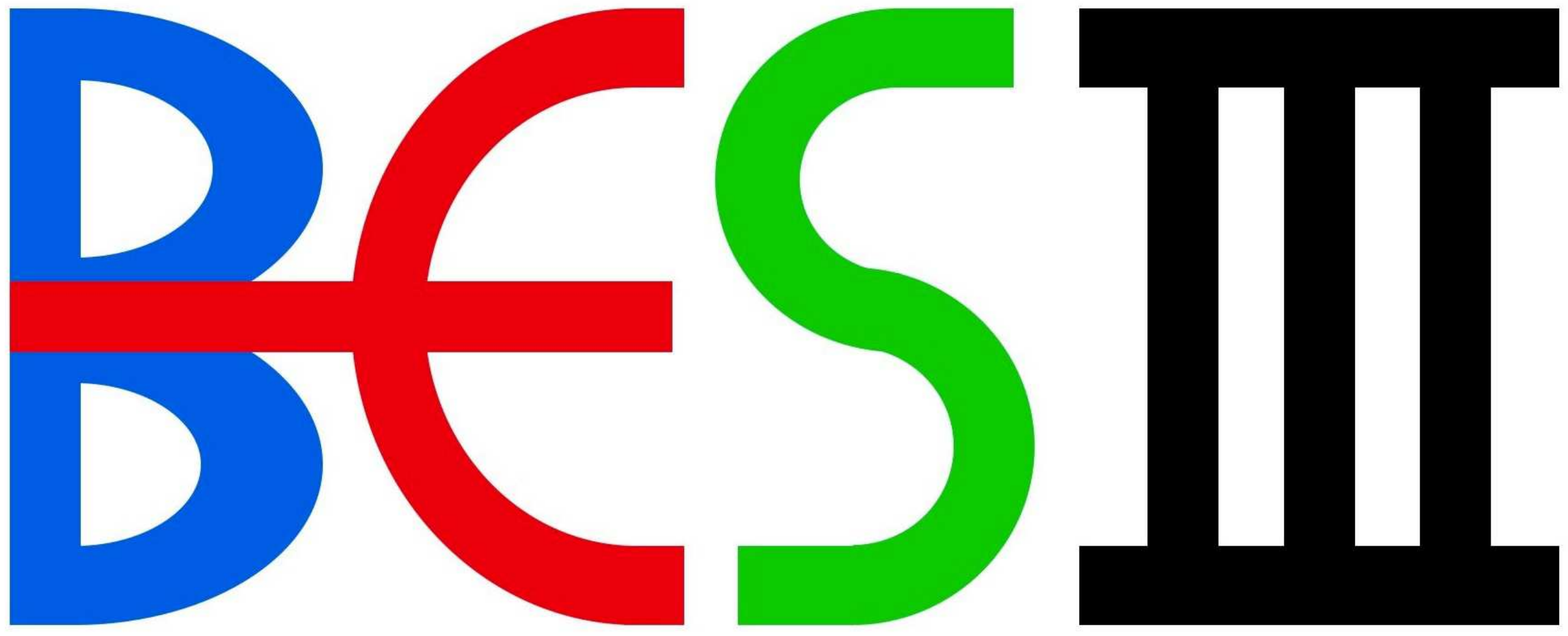}}

\collaboration{The BESIII collaboration}


\emailAdd{besiii-publications@ihep.ac.cn}

\arxivnumber{2209.14564}

\abstract{
    Using $(448.1\pm2.9)\times10^6$ $ \psi(3686)$ events collected with the BESIII detector at the BEPCII collider, the decay $\psi(3686)\to\Sigma^-\bar\Sigma^+$ is observed for the first time with a branching fraction of $(2.82\pm0.04_{\rm stat.}\pm0.08_{\rm syst.})\times10^{-4}$, and the angular parameter $\alpha_{ \Sigma^-}$ is measured to be $0.96\pm0.09_{\rm stat.} \pm 0.03_{\rm syst.}$. 
}

\begin{document}
\maketitle
\flushbottom

\section{Introduction}
Two-body baryonic decays of the charmonium states  $J/\psi$ and $\psi(3686)$, here both denoted by the symbol $\Psi$,  provide an excellent laboratory to study flavour-SU(3) symmetry breaking and test various aspects of quantum chromodynamics (QCD) in the transition region between perturbative and non-perturbative energy regime~\cite{pQCD}.  

The amplitudes of  $\Psi$  decays to different baryon octet pairs are supposed to be the same under the assumption of flavour-SU(3) symmetry. However, branching fractions are not only determined by strong interaction amplitudes, but also by
electromagnetic interactions and interference between the two amplitudes~\cite{Rudaz:1975yu}, although these contributions are much smaller than the expected flavour-SU(3) breaking effects. With a phenomenologically plausible model~\cite{Zhu:2015bha,Ferroli:2019nex, Ferroli:2020mra}, the branching fractions of $\Psi$ decay to baryon octet final states can be described well. 
Perturbative QCD~\cite{Appelquist:1974zd,DeRujula:1974rkb} predicts the partial widths for $\psi(3686)$ decay into an exclusive hadronic state to be proportional to squares of the wave-function, which are well determined from leptonic widths. 
Furthermore, the ratio between the branching fractions of $J/\psi$ and $\psi(3686)$ decays to
the same final states is expected to obey the so-called ``$12\%$ rule''~\cite{Appelquist:1974zd,DeRujula:1974rkb}. 
Although a large fraction of exclusive decay channels follow this rule
approximately, significant violation has been observed in the $\rho
\pi$ channel~\cite{Franklin:1983ve}. The ratio of the branching fraction $\mathcal{B}(\psi(3686) \to \rho \pi)$ to
$\mathcal{B}(J/\psi \to \rho \pi)$ is much smaller than the perturbative QCD prediction, and this is called the
``$\rho\pi$ puzzle''. 
Many explanations~\cite{Ref:review2} of the $\rho\pi$ puzzle have been proposed, including
the $J/\psi$-glueball admixture scheme~\cite{Li:2007ky}, the intrinsic-charm-component scheme~\cite{Brodsky:1997fj}, the sequential-fragmentation model~\cite{Karl:1984en}, the exponential form-factor model~\cite{Chaichian:1988kn}, the $S$-$D$ wave-mixing scheme~\cite{Liu:2004un,Rosner:2001nm}, the final-state interaction scheme and others~\cite{Li:1996yn}.
However, none of these explanations can account for all existing experimental results.
Tests of the $12\%$ rule using the baryonic decay modes are helpful in understanding the $\rho \pi$ puzzle.
Experimentally, the branching fractions of $\Psi$ decay into octet baryon pairs have been well measured, except for $\Sigma^-\bar{\Sigma}^+$\cite{pdg}.

The angular distribution of a baryon pair can be written as $1 + \alpha_{B}\cos^2\theta_{B}$, where $\alpha_B$ is the angular distribution parameter of the baryon, $\theta_{B}$ is a polar angle between the baryon and the positron beam in the centre-of-mass (c.m.) system. 
The value of $\alpha_{B}$ is expected to be 1 due to the helicity conservation rule~\cite{Brodsky:1981kj}. In addition, in the theoretical calculations of $\alpha_{B}$, the masses of quarks and baryons have been considered~\cite{Claudson:1981fj, Carimalo:1985mw}.
Existing theoretical predictions are not consistent with the experimental measurements. 
The values of $\alpha_{B}$ should be the same among  isospin partners, such as $\alpha_{\Sigma^{+}}$ and $\alpha_{\Sigma^{0}}$~\cite{BESIII:2020fqg, BESIII:2017kqw}, $\alpha_{\Xi^{0}}$ and $\alpha_{\Xi^{-}}$~\cite{BESIII:2016nix, BESIII:2016ssr}. There are no significant differences observed experimentally. However, the value of $\alpha_{\Sigma^{-}}$ has not yet been measured.

In this paper, the first observation of the decay $\psi(3686)\to\Sigma^-\bar\Sigma^+$ is reported, where $\Sigma^-$ decays to $n\pi^-$ and $\bar\Sigma^+$ decays to $\bar n\pi^+$. The data samples used in this analysis consist of  $(448.1\pm2.9)\times10^6$ $\psi(3686)$ events~\cite{npsp} collected with the BESIII detector.

\section{The BESIII detector and Monte Carlo simulation}

The BESIII detector~\cite{Ablikim:2009aa} records symmetric $e^+e^-$ collisions 
provided by the BEPCII storage ring~\cite{Yu:IPAC2016-TUYA01}
in the c.m. energy range from 2.0 to 4.94~GeV, with a peak luminosity of $1\times10^{33}$~cm$^{-2}$s$^{-1}$ achieved at $\sqrt{s}=3.77$ GeV.
BESIII has collected large data samples in this energy region~\cite{Ablikim:2019hff}. The cylindrical core of the BESIII detector covers 93\% of the full solid angle and consists of a helium-based
 multilayer drift chamber~(MDC), a plastic scintillator time-of-flight
system~(TOF), and a CsI(Tl) electromagnetic calorimeter~(EMC),
which are all enclosed in a superconducting solenoidal magnet
providing a 1.0~T (0.9~T in
2012) magnetic field. The solenoid is supported by an
octagonal flux-return yoke with resistive plate counter muon
identification modules interleaved with steel. 
The charged-particle momentum resolution at $1~{\rm GeV}/c$ is
$0.5\%$, and the  d$E$/d$x$ resolution is $6\%$ for electrons
from Bhabha scattering. The EMC measures photon energies with a
resolution of $2.5\%$ ($5\%$) at $1$~GeV in the barrel (end-cap)
region. The time resolution in the TOF barrel region is 68~ps, while
that in the end-cap region is 110~ps.

Monte Carlo~(MC) simulated events are used to determine the detection
efficiency, optimize selection criteria, and study possible
backgrounds.  
Simulated data samples produced with a {\sc
geant4}-based~\cite{geant4} package, which
includes the geometric description of the BESIII detector and the
detector response, are used to determine detection efficiencies
and to estimate backgrounds. The simulation models the beam-energy spread and initial-state radiation (ISR) in the $e^+e^-$
annihilations with the generator {\sc
kkmc}~\cite{KKMC}. The inclusive MC sample $\psi(3686)$ includes the production of the
$\psi(3686)$ resonance, the ISR production of the $J/\psi$, and
the continuum processes incorporated in {\sc
kkmc}. The known decay modes are modeled with 
BesEvtGen~\cite{evtgen}
using branching fractions taken from the
Particle Data Group (PDG)~\cite{pdg}, and the remaining unknown charmonium decays
are modelled with {\sc lundcharm}~\cite{lundcharm}. Final-state radiation
from charged final-state particles is incorporated using the {\sc
photos} package~\cite{photos}. The differential cross
section of the signal process~($\psi(3686)\to\Sigma^-\bar\Sigma^+$, $\Sigma^-\to n \pi^-$, $\bar\Sigma^+\to \bar n \pi^+$) is expressed with respect to five observables
$\boldsymbol{\xi}= ( \theta_{\Sigma^{-}}, \theta_{n}, \phi_{n},
\theta_{\overline{n}},
\phi_{\overline{n}})$, and includes four parameters $\alpha_{\Sigma^{-}}$, $\Delta\Phi$, $\alpha_{-}$ and $\alpha_{+}$~\cite{Faldt:2017kgy}. 
Here, $\theta_{\Sigma^{-}}$
is the polar angle between the $\Sigma^{-}$ and the positron beam in
the reaction c.m. frame, $\theta_{n}, \phi_{n}$ and
$\theta_{\overline{n}}, \phi_{\overline{n}}$ are the polar and
azimuthal angles of the neutron and anti-neutron measured in the rest
frames of their corresponding parent particles. 
The value of $\alpha_{\Sigma^{-}}$ is determined in this analysis, and $\Delta\Phi$ is set to be 0 by assuming no polarization.
The decay asymmetry parameters $\alpha_{-}$ and $\alpha_{+}$ in the differential cross sections are fixed to $-0.068$ and $0.068$ using the PDG~\cite{pdg} values, where $\alpha_{-}$ and $\alpha_{+}$ are used to describe the non-leptonic decays of $\Sigma^-\to n\pi^-$ and $\bar\Sigma^+\to \bar n\pi^+$~\cite{Lee:1957qs}. The uncertainties in the values of these parameters are considered when assigning systematic uncertainties.

\section{Event selection}

The final state of the signal process is $n\pi^-\bar{n}\pi^+$. Event candidates are required to have two well-reconstructed charged tracks with zero net charge, and one anti-neutron. In order to keep the selection efficiency high there is no attempt made to reconstruct the neutron.
Charged tracks detected in the MDC are required to be within a polar angle ($\theta$) range of $|\!\cos\theta|<0.93$ and $\theta$ is defined with respect to the $z$ axis, which is along the symmetry axis of the MDC. For each charged track, the distance of closest approach to the interaction point (IP) must be less than 30\,cm along the $z$ axis, and less than 10\,cm in the transverse plane.
Particle identification~(PID) for charged tracks combines measurements of the energy deposited in the MDC~(d$E$/d$x$) and the flight time in the TOF to form likelihoods $\mathcal{L}(h)~(h=p,K,\pi)$ for each hadron $h$ hypothesis.
Two pions are identified with the requirements that $\mathcal{L}(\pi)>\mathcal{L}(K)$ and $\mathcal{L}(\pi)>\mathcal{L}(p)$.

The anti-neutron candidates are identified using showers in the EMC. The deposited energy of each shower must be more than 600~MeV both in the barrel region ($|\!\cos \theta|< 0.80$) and in the end-cap region ($0.86 <|\!\cos \theta|< 0.92$).  
To exclude showers that originate from charged tracks, the angle subtended by the EMC shower and the position of the closest charged track at the EMC must be greater than 10 degrees as measured from the IP.
The second moment $\sum_i E_i r_i^2 /\sum_i E_i$, where $E_i$ is the energy deposition in the $i^{\rm th}$ crystal and $r_i$ is the radial distance of the $i^{\rm th}$ crystal from the cluster centre, is required to be larger than 20, to suppress the photon background misidentified as anti-neutrons.
To suppress electronic noise and showers unrelated to the event, the difference between the EMC time and the event start time is required to be within [0, 700]\,ns. If the number of anti-neutron candidates in an event is more than one, the most energetic candidate in the EMC is selected.

A kinematic fit is performed to the decay $\psi(3686)\to n\pi^-\bar{n}\pi^+$ with the constraints provided by four-momentum conservation and by the invariant mass $M_{\bar{n} \pi^{+}}$ equal to the known $\bar{\Sigma}^{+}$ mass. Since the anti-neutron could annihilate with the materials in the EMC, the polar and azimuthal angles of anti-neutron are used in kinematic fit, while the anti-neutron deposited energy is left free. Considering that the neutron is hardly to be detected, the neutron three-momentum components are left as free parameters in the fit. The $\chi^2$  of the  kinematic fit  is required to be smaller than 50, which is a value optimized by using the figure-of-merit $S/\sqrt{S+B}$, where $S$ is the number of signal MC events and $B$ is the number of the estimated background events. To suppress background from $\psi(3686) \rightarrow \pi^+ \pi^- J/\psi$ decays, with $J/\psi \rightarrow n \bar{n}$, the recoil mass of the $\pi^{+} \pi^{-}$ pair is required to be less than 2.9 GeV/$c^{2}$.

An inclusive MC sample of 506 million $\psi(3686)$ events is used to study possible background channels, with a generic event-type analysis tool, TopoAna~\cite{ref:topo}. The potential sources of  peaking background are found to be $\psi(3686)\to\gamma\chi_{cJ}(\chi_{cJ}\to\Sigma^-\bar\Sigma^+)~(J=0,1,2)$ and $\psi(3686)\to\gamma\eta_c(\eta_c\to\Sigma^-\bar\Sigma^+)$, and $\psi(3686)\to\pi^0\Sigma^-\bar\Sigma^+$. 
To estimate the sizes and distributions of these background processes, samples of 100 million events are generated for each channel.  In these, the decay processes $\psi(3686)\to\gamma\chi_{cJ}(\chi_{cJ}\to\Sigma^-\bar\Sigma^+)$, $\psi(3686)\to\gamma\eta_c(\eta_c\to\Sigma^-\bar\Sigma^+)$, and $\psi(3686)\to\pi^0\Sigma^-\bar\Sigma^+$ are generated with the P2GCJ (J=0,1,2), JPE, and phase-space models. When accounting for the branching fractions~\cite{pdg} and detection efficiencies of these decays,   the numbers of background events passing the selection in the data sample are predicted to be  $562\pm86$ for $\psi(3686)\to\gamma\chi_{cJ}(\chi_{cJ}\to\Sigma^-\bar\Sigma^+)$ and $5\pm1$ for $\psi(3686)\to\gamma\eta_c(\eta_c\to\Sigma^-\bar\Sigma^+)$. 
The contribution of  $\psi(3686)\to\pi^0\Sigma^-\bar\Sigma^+$ decays is negligible.

An off-resonance data sample taken at the c.m. energy of 3.65 GeV is used to estimate the non-$\psi(3686)$ background.
The size of this contribution,  $N_{\mathrm{non-}\psi(3686)}$, is determined according to the formula: $N_{\mathrm{non-}\psi(3686)}=N_{\mathrm{cont}}^{\mathrm{obs}}\cdot\frac{L_{\psi(3686)}}{L_{\mathrm{cont}}}\cdot\frac{s_{\mathrm{cont}}}{s_{\mathrm{\psi(3686)}}}\cdot\frac{\varepsilon_{\mathrm{cont}}}{\varepsilon_{\mathrm{\psi(3686)}}}=92\pm53$ events, where $N_{\mathrm{cont}}^{\mathrm{obs}}=7\pm4$ is the number of surviving events under identical selection criteria when applied to the 3.65 GeV sample, $s_{\mathrm{cont}}$ and $s_{\mathrm{\psi(3686)}}$ are the  squares of the c.m. energies at 3.65 GeV and 3.686 GeV, $\varepsilon_{\mathrm{cont}}=5.96\%$ and $\varepsilon_{\mathrm{\psi(3686)}}=5.26\%$ are the selection efficiencies at 3.65 GeV and 3.686 GeV, and $L_{\mathrm{cont}} = 44\,{\rm  pb}^{-1}$ and $L_{\mathrm{\psi(3686)}}=668.55\,{\rm pb}^{-1}$ and are the integrated luminosities at 3.65 GeV and 3.686 GeV, respectively.

\section{Measurement of the branching fraction}
\begin{figure}[hbtp]
\centering
\includegraphics[trim=0 50 0 50,clip,width=0.8\textwidth]{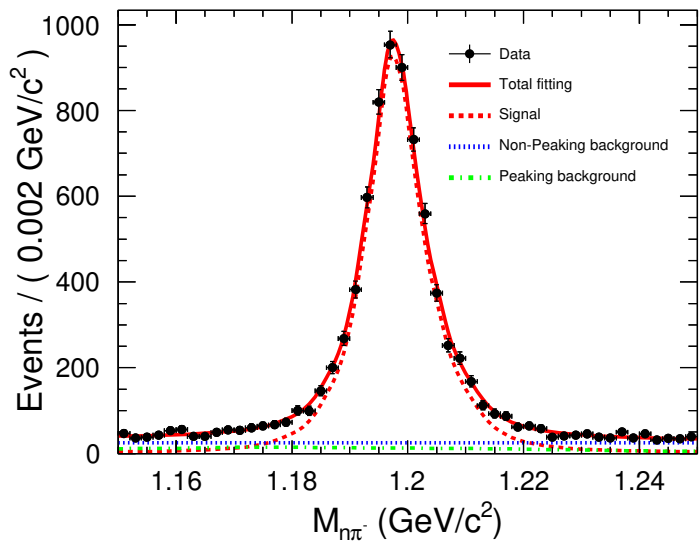}
\caption{ The distribution of $M_{n\pi^-}$ for the process $\psi(3686)$ decay to final state $n \bar n \pi^+\pi^-$. The black dots with error bars are the data, the red solid line is the total fit function, the red dashed line is the signal function, the blue dotted line is the non-peaking background function, and the green dash-dotted line is the peaking background function.   }
\label{fig:fitting_result}
\end{figure}

The $\Sigma^-$ candidate is reconstructed from the $\pi^-$ and the missing neutron. To determine the number of signal events, an unbinned maximum likelihood fit is performed to the distribution of the invariant mass $n\pi^-$ ($M_{n\pi^{-}}$) in the region of [1.15, 1.25] GeV/$c^{2}$. The signal is described by the shape found in the MC simulations, convoluted with a Gaussian function which accommodates any difference in mass resolution between data and MC simulations. The peaking background is described with the shapes of the MC-simulated exclusive background channels, and the corresponding numbers of events are fixed to the estimated values. The non-peaking background is described with a first-order polynomial function since the distribution contributed by total possible backgrounds is observed to be almost uniform by studying the inclusive MC sample of 506 million $\psi(3686)$. 
Figure~\ref{fig:fitting_result} shows the fit of the $n\pi^{-}$ mass distribution.
The $\chi^2/{\rm ndf}$  of the fit is 35.64/44, where $\rm ndf$ is the number of degrees of freedom. 

The branching fraction is calculated according to
\begin{equation}
 \begin{aligned}
&Br=\frac{N_{\mathrm{cand}} - N_{\mathrm{backg-}\psi(3686)}-N_{\mathrm{non-}\psi(3686)}}{\varepsilon\times \prod{Br_i}\times N_{\mathrm{tot}}} , \\
 \end{aligned}
 \end{equation}
 where $N_{\mathrm{cand}}$ is the number of events selected by the kinematic fit, $N_{\mathrm{backg-}\psi(3686)}$ is the number of $\psi(3686)$ background events including non-peaking background events $N_{\mathrm{non-peakbackg-}\psi(3686)}$ and peaking background events $N_{\mathrm{peakbackg-}\psi(3686)}$, $N_{\mathrm{non-}\psi(3686)}$ is the number of non-$\psi(3686)$ events, $\prod{Br_i}$ is the product of the branching fractions of the intermediate states, and $N_{\mathrm{tot}}$ is the number of total $\psi(3686)$ events~\cite{npsp}.
The detection efficiency $\varepsilon$ is estimated from the signal MC simulations. 
Differences in detection efficiency between data and MC simulations is accounted for using control samples of $J/\psi\to p\bar{p}\pi^+\pi^-$, $J/\psi\to p\bar{n}\pi^-$, and $\psi(3686)\to p\bar{n}\pi^-$ decays. Here, in order to study the difference of anti-neutron  efficiency from EMC and kinematic fit, the anti-neutron efficiency ratios between data and MC simulations are determined with different anti-neutron momentum and polar-angle regions. Besides, the efficiency difference, the polar and azimuth angles of anti-neutron, and their error matrices have been corrected based on the data-driven method~\cite{Liu:2021rrx}. The $\pi^+$ and $\pi^-$ efficiency ratios are also determined using the same method. 
The branching fraction is calculated to be $(2.82\pm 0.04) \times10^{-4}$, where the uncertainty is statistical only. The corresponding number of events $N_{\mathrm{cand}}$ selected by the kinematic fit, non-$\psi(3686)$ background events $N_{\mathrm{non-}\psi(3686)}$,  non-peaking background events $N_{\mathrm{non-peakbackg-}\psi(3686)}$ from $\psi(3686)$ decay, peaking background events $N_{\mathrm{peakbackg-}\psi(3686)}$ from $\psi(3686)$ decay, and detection efficiency $\varepsilon$ after correcting for data-MC differences are listed in Tab.~\ref{our br}.

 \begin{table}[hbtp]
 \centering
\caption{The number of total events selected by the kinematic fit, the number of non-$\psi(3686)$ background events, the number of non-peaking background events from $\psi(3686)$ decay, the number of peaking background events from $\psi(3686)$ decay, and the  detection efficiency.  } 
\label{our br}
\begin{tabular}{ccccccc}
\hline \hline 
Channel &$N_{\mathrm{cand}}$ &$N_{\mathrm{non-}\psi(3686)}$ &$N_{\mathrm{non-peakbackg-}\psi(3686)}$&$N_{\mathrm{peakbackg-}\psi(3686)}$&$\varepsilon$ \\\hline 
$\psi(3686)\to\Sigma^-\bar\Sigma^+$&$8536$ & $92\pm53$& $1253\pm62$& $562\pm86$ &$5.26\%$ \\
 \hline \hline
\end{tabular}
\end{table}

\section{Measurement of the angular distribution parameter}
\begin{figure}[hbtp]
\centering
\includegraphics[trim=0 50 0 50,clip,width=0.8\textwidth]{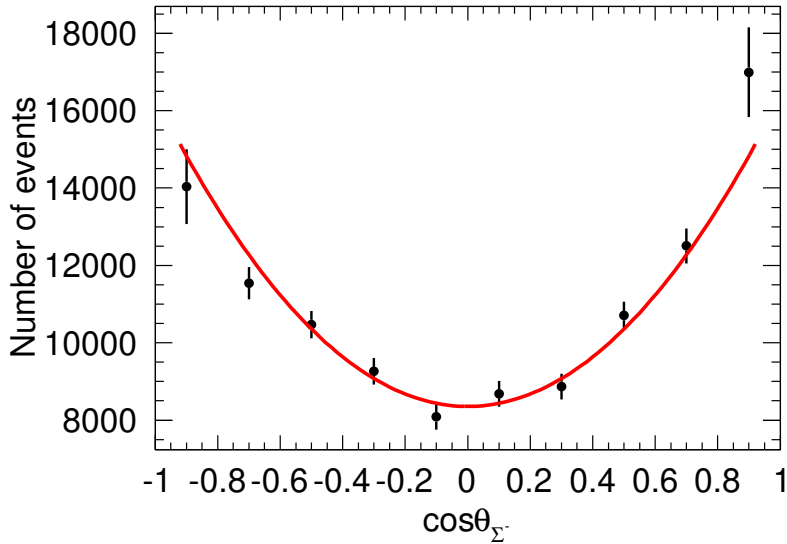}
\caption{ The angular distribution for the signal process $\psi(3686)\to\Sigma^-\bar\Sigma^+$, $\Sigma^-\to n \pi^-$, $\bar\Sigma^+\to \bar n \pi^+$. The black dots with error bars indicate the signal yields after efficiency correction, and the red curve represents the fit function. }
\label{fig:angular distribution}
\end{figure}

 To determine the value of the angular parameter $\alpha_{\Sigma^-}$, a least-squares fit  is performed to the $\cos\theta_{\Sigma^-}$ distribution in the range of [$-$1, 1]. 
The numbers of signal events   are determined in ten equally sized  intervals of $\cos\theta_{\Sigma^-}$ with the same method as used in the branching fraction measurement. The detection efficiency in each interval is determined with  MC simulations, which is then corrected to account for data-MC differences.  The $\cos\theta_{\Sigma^-}$ distribution after efficiency correction is shown in Fig.~\ref{fig:angular distribution}.  Superimposed is the result of a fit to the function   $1 + \alpha_{\Sigma^-} \cos^2\theta_{\Sigma^-}$.
The parameter $\alpha_{\Sigma^-}$ is measured to be $0.96\pm0.09$, where the uncertainty is statistical, and its lower limit is determined to be larger than 0.835 at 90\% confidence level. The  $\chi^2/{\rm ndf}$ of the fit is $13.52/8$, where $\rm ndf$ is the number of degrees of freedom. 

\section{Systematic uncertainties}
To estimate the systematic uncertainties in the measurement of the branching fraction, we consider the differences of the detection efficiency and resolution between data and MC simulations, the uncertainty associated with the generator models, the  background estimations and other sources. 
An overview of all the systematic uncertainties on the branching fraction measurement is given in Tab.~\ref{table6}.

\begin{table}[htbp]
 \centering
\caption{Systematic uncertainties on the branching fraction measurement (\%). }
\begin{tabular}{lc}\hline\hline
 \multicolumn{1}{c}{ Source } &  \multicolumn{1}{c}{ Uncertainty~(\%)} \\
\hline
MC efficiency correction                                                   & 1.9 \\
Decay parameter                                                               & 1.2  \\
QED peaking-background estimation                                 & 0.8   \\
Non-peaking background estimation                                  & 0.4   \\
Peaking-background estimation                                          & 0.6   \\
Kinematic fitting                                                                   & 0.2   \\
Total number of $\psi(3686)$                                               & 0.7   \\ \hline
Total                                                                                     & 2.6   \\\hline
\hline
\end{tabular}
\label{table6}
\end{table}

The tracking and PID efficiency in MC simulations is corrected in bins of transverse momentum and polar angle to agree with that measured in data.  The uncertainty on these corrections, derived from the control channels and averaged over bins, is assigned as a systematic uncertainty on the branching fraction.   For charged tracks the study is performed with a control sample of  $J/\psi\to p\bar p \pi^+\pi^-$ events, and the relative uncertainty is found to be 0.2\%.   The relative uncertainty for the reconstruction of the anti-neutron is set with control samples of  $J/\psi\to p\bar{n}\pi^-$ and $\psi(3686)\to p\bar{n}\pi^-$ events and found to be 1.9\%.  Hence, the total systematic uncertainty associated with the MC efficiency correction is 1.9\%.

In the signal generator model~\cite{Faldt:2017kgy}, the values of $\alpha_{-}$ and $\alpha_{+}$ are set to be $-0.068$ and $0.068$ for $\Sigma^{-}$ and $\bar{\Sigma}^{+}$ respectively. Furthermore, we assume that there is no polarization by setting $\Delta\Phi$ to 0. To evaluate the systematic uncertainty associated with these assumptions, we vary $\alpha_{-}$ and $\alpha_{+}$ by one standard deviation (0.008), and change $\Delta\Phi$ to be $-\pi$ or $+\pi$. We compare the efficiencies after these variations with the baseline efficiency, and take the maximum difference, 1.2\%,  as the corresponding systematic uncertainty.

Possible systematic effects due to the requirement of $M_{\rm rec}(\pi^+\pi^-) < 2.9$ GeV/$c^{2}$ are investigated by varying the selection criteria between 2.80 and 2.91 GeV/$c^{2}$ in steps  of 1 MeV/$c^{2}$.  The variations observed are compatible with statistical fluctuations and thus no uncertainty is assigned associated with this requirement~\cite{Barlow:2002yb}.  

The uncertainty associated with non-peaking background is estimated by changing the order of polynomial function used to describe this background. The difference of 0.4\% with respect to the baseline configuration is taken as the systematic uncertainty arising from this source.  

The uncertainty associated with the number of non-$\psi(3686)$ background events is assigned by varying the sizes of these backgrounds by one standard deviation, giving contributions of $0.8\%$.

The uncertainty associated with the peaking background is assigned by varying the sizes of these backgrounds by one standard deviation. Besides, the $\chi_{c,J}\to\Sigma^-\bar\Sigma^+$ decays are generated with the ANGSAM model, with helicity angles $\theta$ of the $\Sigma^-$ satisfying the angular distribution $1+\alpha\times\cos^{2}\theta$, where $\alpha$ is the angular distribution parameter of the baryon. Two extreme cases in the analysis are performed to consider the expected detection efficiency and mass distribution, namely with $\alpha=1$ and $-1$. The maximum difference $0.6\%$ is taken as systematic uncertainty.

To estimate the size of any potential bias arising from the  kinematic fit, we obtain the $\chi^2$ distributions with the track correction method for the helix parameters that are corrected to reduce the differences between data and MC simulations~\cite{BESIII:2012mpj}. Besides, the polar and azimuth angles and error matrix of anti-neutron in kinematic fit have also been corrected~\cite{Liu:2021rrx}. 
Compared with the baseline value, the difference of 0.2\% is taken as the systematic uncertainty.
An uncertainty of 0.7\% is assigned to reflect the knowledge of the  number of $\psi(3686)$ events in the sample, which  is measured from inclusive hadronic decays, as described in Ref.~\cite{npsp}.

The main sources of systematic uncertainty on baryonic angular distribution measurement are associated with knowledge of the signal yields, the efficiency correction, and the fitting process. An overview of all the systematic uncertainties is given in Tab.~\ref{table7}.  

\begin{table}[htbp]
 \centering
\caption{Systematic uncertainties of angular-distribution measurement (\%). }
\begin{tabular}{lc}\hline\hline
 \multicolumn{1}{c}{ Source } &  \multicolumn{1}{c}{ Uncertainty(\%)} \\
\hline
MC efficiency correction                                                              & 0.5 \\
QED peaking background estimation                                        & negligible   \\
Non-peaking background estimation                                         & 1.4   \\
Peaking background estimation                                                 & 1.7   \\
Kinematic fitting                                                                           & 0.3   \\
Number of bins                                                                           & 0.4  \\
Fitting $\cos\theta_{\Sigma^{-}}$ range                                                            & 1.9   \\ \hline
Total                                                                                             & 3.0   \\\hline
\hline
\end{tabular}
\label{table7}
\end{table}

In the angular distribution measurement,  the number of signal events in each bin is determined by the same method as for the branching fraction measurement. The uncertainties on this yield determination are associated with the MC efficiency correction, background estimation and kinematic-fitting requirement. These uncertainties are estimated with the same method as for the branching fraction.   In doing this, we consider the correlations between the measurements in each bin.
We then re-perform the fit to the  angular distribution and take the difference with respect to baseline value as the systematic uncertainty for each contribution.
The uncertainties associated with the $\alpha_{\Sigma^{-}}$ fit itself are estimated 
by varying the fitting range  in $\cos\theta_{\Sigma^{-}}$ from [$-1.0$, 1.0] to [$-0.8$, 0.8], and also changing the number of bins from ten to eight.  In both cases the changes in result are assigned as contributions to the uncertainties.    The total systematic uncertainty on $\alpha_{\Sigma^{-}}$ is 0.029.

\section{Summary}
In summary, based on the $(448.1\pm2.9)\times 10^{6}$ $\psi(3686)$ events collected at BESIII detector, the branching fraction and angular parameter, $\alpha_{\Sigma^-} $, of $\psi(3686) \to\Sigma^-\bar\Sigma^+$ decays are measured for the first time.  The measurements yield  $(2.82 \pm 0.04_{\rm stat.} \pm 0.08_{\rm syst.})\times10^{-4}$ for the branching fraction and $\alpha_{\Sigma^-} = 0.96 \pm 0.09_{\rm stat.} \pm 0.03_{\rm syst.}$. Table~\ref{table:summary} summarizes measurements of the angular parameter and branching fractions for $\psi(3686)\to\Sigma^+\bar\Sigma^-$, $\Sigma^0\bar\Sigma^0$, and $\Sigma^-\bar\Sigma^+$ channels, and predicted values for the branching fractions.  The measured branching fraction is around 2.3$\sigma$ above the theoretical prediction value $(2.46 \pm 0.13)\times10^{-4}$~\cite{Ferroli:2020mra}. The contributions from strong, electromagnetic, and their interference may explain that, although there are some discrepancy between them. Considering the experimental uncertainties, they are consistent within 3$\sigma$. 
 There are  significant differences between the value of  $\alpha_{\Sigma^-}$  and  those of its  isospin partners $\alpha_{\Sigma^{+}}$ and $\alpha_{\Sigma^{0}}$ , which are worthy of further investigation.  Finally,  it is noted that the analysis method pursued here can also be used to measure the branching fraction of $J/\psi\to\Sigma^-\bar\Sigma^+$, which in combination with the result reported in this paper will  provide an opportunity to further test the ``$12\%$ rule'' in charmonium decays. 

 \begin{table}[hbtp]
 \centering
\caption{Summary of the measured angular parameters and branching fractions of $\psi(3686)\to\Sigma^+\bar\Sigma^-$, $\Sigma^0\bar\Sigma^0$, and $\Sigma^-\bar\Sigma^+$, together with theoretical predictions of the branching fractions.} 
\label{table:summary}
\small
\begin{tabular}{cccc}
\hline \hline 
Decay mode&Br($\times 10^{-4}$)&Angular parameter $\alpha_{B}$&Br prediction($\times 10^{-4}$)~\cite{Ferroli:2020mra}\\\hline 
$\psi(3686)\to\Sigma^+\bar\Sigma^-$&$2.52\pm 0.04 \pm 0.09$~\cite{BESIII:2021wkr}&$0.682 \pm 0.030 \pm 0.011$~\cite{BESIII:2020fqg}&$2.29 \pm 0.15$\\\hline 
$\psi(3686)\to\Sigma^0\bar\Sigma^0$&$2.44 \pm 0.03 \pm 0.11 $~\cite{BESIII:2017kqw}&$0.71 \pm 0.11 \pm 0.04 $~\cite{BESIII:2017kqw}&$2.37 \pm 0.09$ \\\hline 
$\psi(3686)\to\Sigma^-\bar\Sigma^+$&$2.82 \pm 0.04 \pm 0.08$&$0.96 \pm 0.09 \pm 0.03$&$2.46 \pm 0.13$\\
 \hline \hline
\end{tabular}
\end{table}

\acknowledgments
The BESIII collaboration thanks the staff of BEPCII and the IHEP computing center for their strong support. This work is supported in part by National Key R\&D Program of China under Contracts Nos. 2020YFA0406300, 2020YFA0406400; National Natural Science Foundation of China (NSFC) under Contracts Nos. 11635010, 11735014, 11835012, 11935015, 11935016, 11935018, 11961141012, 12022510, 12025502, 12035009, 12035013, 12192260, 12192261, 12192262, 12192263, 12192264, 12192265; the Chinese Academy of Sciences (CAS) Large-Scale Scientific Facility Program; Joint Large-Scale Scientific Facility Funds of the NSFC and CAS under Contract No. U1832207; the CAS Center for Excellence in Particle Physics (CCEPP); 100 Talents Program of CAS; The Institute of Nuclear and Particle Physics (INPAC) and Shanghai Key Laboratory for Particle Physics and Cosmology; Sponsored by Shanghai Pujiang Program(20PJ1401700); ERC under Contract No. 758462; European Union's Horizon 2020 research and innovation programme under Marie Sklodowska-Curie grant agreement under Contract No. 894790; German Research Foundation DFG under Contracts Nos. 443159800, Collaborative Research Center CRC 1044, GRK 2149; Istituto Nazionale di Fisica Nucleare, Italy; Ministry of Development of Turkey under Contract No. DPT2006K-120470; National Science and Technology fund; National Science Research and Innovation Fund (NSRF) via the Program Management Unit for Human Resources \& Institutional Development, Research and Innovation under Contract No. B16F640076; STFC (United Kingdom); Suranaree University of Technology (SUT), Thailand Science Research and Innovation (TSRI), and National Science Research and Innovation Fund (NSRF) under Contract No. 160355; The Royal Society, UK under Contracts Nos. DH140054, DH160214; The Swedish Research Council; U. S. Department of Energy under Contract No. DE-FG02-05ER41374.

\clearpage

\section*{The BESIII Collaboration}
\addcontentsline{toc}{section}{The BESIII Collaboration}
\begin{small}
M.~Ablikim$^{1}$, M.~N.~Achasov$^{11,b}$, P.~Adlarson$^{70}$, M.~Albrecht$^{4}$, R.~Aliberti$^{31}$, A.~Amoroso$^{69A,69C}$, M.~R.~An$^{35}$, Q.~An$^{66,53}$, X.~H.~Bai$^{61}$, Y.~Bai$^{52}$
% ...
\\
\\
{\it
$^{1}$ Institute of High Energy Physics, Beijing 100049, People's Republic of China\\
% ...
$^{4}$ Bochum Ruhr-University, D-44780 Bochum, Germany\\
% ...
$^{11}$ G.I. Budker Institute of Nuclear Physics SB RAS (BINP), Novosibirsk 630090, Russia\\
% ...
$^{31}$ Johannes Gutenberg University of Mainz, Johann-Joachim-Becher-Weg 45, D-55099 Mainz, Germany\\
% ...
$^{35}$ Liaoning Normal University, Dalian 116029, People's Republic of China\\
% ...
$^{52}$ Southeast University, Nanjing 211100, People's Republic of China\\
$^{53}$ State Key Laboratory of Particle Detection and Electronics, Beijing 100049, Hefei 230026, People's Republic of China\\
% ...
$^{61}$ University of Jinan, Jinan 250022, People's Republic of China\\
% ...
$^{66}$ University of Science and Technology of China, Hefei 230026, People's Republic of China\\
% ...
$^{69}$ University of Turin and INFN, (A)University of Turin, I-10125, Turin, Italy; (B)University of Eastern Piedmont, I-15121, Alessandria, Italy; (C)INFN, I-10125, Turin, Italy\\
$^{70}$ Uppsala University, Box 516, SE-75120 Uppsala, Sweden\\
% ...
$^{b}$ Also at the Novosibirsk State University, Novosibirsk, 630090, Russia\\
% ...
}
\end{small}

\end{document}